\documentclass[aps,twocolumn,prl]{revtex4}

\usepackage{graphicx}
\usepackage{epstopdf}
\usepackage{epsfig}
\usepackage{threeparttable}
\usepackage{times}

\begin{document}

\title{Cooperativity, sensitivity and noise in biochemical signaling}

\author{William Bialek$^a$ and Sima Setayeshgar$^b$}

\affiliation{$^a$Joseph Henry Laboratories of Physics and the Lewis--Sigler Institute for Integrative Genomics,
Princeton University, Princeton, New Jersey 08544,
$^b$Department of Physics, Indiana University, Bloomington, Indiana 47405}

\begin{abstract}
Cooperative interactions among the binding of multiple signaling molecules  is a common mechanism for enhancing the sensitivity of biological signaling systems.  It is widely assumed that this increase in sensitivity of the mean response implies the ability to detect smaller signals.  We show that, quite generally, there is a component of the noise in such systems that can be traced to the random arrival of the signaling molecules at their receptor sites, and this diffusive noise is not reduced by cooperativity.  Cooperativity makes it easier for real systems to reach this physical limit, but cannot reduce the limit itself.
\end{abstract}

\date{\today}

\maketitle 
 
In biological systems, many signals are carried by changes in the concentration of various  molecules, from small molecules such as cyclic nucleotides to proteins such as the transcription factors that regulate gene expression.   A striking feature of these systems is their sensitivity:  concentrations can be quite small in absolute terms, and small fractional changes in concentration can have a significant effect.   Thus, many transcription factors achieve their half maximal effect at nanoMolar concentrations \cite{nanoMolar}, corresponding to just a handful of molecules per cubic micron, 
changes in the concentration of a particular phosphorylated protein by less than $\sim 10\%$ can produce reliable differences in the output of the bacterial flagellar motor \cite{cluzel+al_00}, and so on.  A natural question about the physics of these systems concerns the origins of this sensitivity and its ultimate limits.

A central theme in much of biological regulation is the idea of cooperativity or allostery \cite{mwc,knf,perutz_90}:  binding of multiple ligands at different sites on a single protein molecule can interact so that the mean occupancy of these sites has a very steep dependence on ligand concentration.  Indeed,  conventional models predict that cooperative effects among a large number of sites can produce arbitrarily high sensitivity.  But  there is a difference between sensitivity in the mean response and a reliable response to small changes---there are limits to measurement which cannot be evaded just by having an  amplifier with a higher gain.  Here we consider the ``noise floor'' for detection of small concentration changes by a receptor that exhibits cooperativity, or more generally multiple internal states in response to binding of multiple ligands.  We find that while cooperativity can reduce the effective noise level against which small signals must be discriminated, there is a physical limit set by the size of the receptor, the diffusion constant of the signaling molecule and its absolute concentration.  Thus cooperativity can make it easier to approach this limit (as with high gain amplifiers), but cooperative systems do not have a fundamentally lower noise floor than non--cooperative systems.

The classical discussion of noise in biochemical signaling is by Berg and Purcell \cite{BP}. 
Their arguments were heuristic, suggesting that one should think of receptors with size $\ell$ as counting molecules in a volume $\sim \ell^3$.  The Poisson statistics of counting then set a limit to precision which can be reduced by averaging over time, but one must wait a time $\sim \ell^2/D$ for the volume to be cleared by diffusion before making a genuinely independent measurement.  The end result is that if one averages for a time $\tau_{\rm avg}$ then one can detect fractional changes in concentration 
\begin{equation}
{{\Delta c}\over {\bar c}} \sim {1\over \sqrt{D\ell{\bar c}\tau_{\rm avg}}} .
\label{bp}
\end{equation}
Bacterial chemotaxis actually seems to operate close to this limit \cite{BP}.  The discussion of noise  has been invigorated by experiments that measure directly the noise in the regulation of gene expression  \cite{gene_noise_expt}, and much  theoretical work is focused on connecting these data to specific kinetic models   \cite{gene_noise_theory}.  Recently we have argued that even for these more complex systems, there is a limit analogous to that derived by Berg and Purcell, which should be independent of the (often unknown) kinetic details \cite{pnas}.  Our argument  was based on the idea that binding of signaling molecules to their receptor sites usually is an equilibrium process, so that fluctuations in the occupancy of the sites becomes a form of thermal noise that can be analyzed using the fluctuation--dissipation theorem.  The first analysis considered multiple binding sites, in particular allowing us to show how correlations among site occupancies leads to behavior that approximates a single larger receptor, but we did not consider interactions among the sites.  

To analyze the case of cooperative interactions we begin with a very general scenario, and then show how this general analysis works out for a specific model. Consider a receptor molecule or complex that has many states labeled by an index ${\rm i}$; the population of each state is $p_{\rm i}$.  In the state $\rm i$, $n_{\rm i}$ signaling molecules are bound, so that the free energy of this state is $F_{\rm i} = E_{\rm i} - n_{\rm i} \mu$, where $\mu$ is the chemical potential of the signaling molecule measured at the location of the receptor.  As usual for a solution at low concentration, $\mu = k_B T \ln c$, where $c$ is the concentration in appropriate units.  The chemical kinetics of this system (which we don't specify in detail) determine some linear response of the populations to changes in the free energies.  In the frequency domain, \begin{eqnarray}
\delta {\tilde p}_{\rm i} (\omega ) &=& \sum_{\rm j} \chi_{\rm ij}  (\omega) \delta {\tilde F}_{\rm j}(\omega) \\
&=& 
\sum_{\rm j} \chi_{\rm ij}  (\omega)\left[ \delta{\tilde E}_{\rm j} (\omega ) - n_{\rm j} k_B T 
{{\delta{\tilde c}({\bf x}_0 ; \omega)}\over {\bar c}}
\right],
\label{chi2}
\end{eqnarray}
where we assume that all binding and unbinding of the signaling molecules can be considered to occur at a single location ${\bf x}_0$ of the receptor.

The susceptibility $\chi_{\rm ij} (\omega)$ can encode  complex interactions among different binding sites, since it describes all possible states of the receptor complex.  In particular, if there are two binding sites which can be empty or occupied, there are four possible states, which we can label ${\rm i} \equiv \{00, 01, 10, 11\}$.  Positive cooperativity in this simplest of models is the statement that the energy decrease on binding the second ligand is greater than on binding the first ligand,
$E_{11} - E_{10} < E_{10} - E_{00} $.
More complex  (and realistic) models ascribe the cooperativity not to direct interaction between the binding sites but rather to an interaction between the binding and some other conformational degrees of freedom in the receptor, as in the Monod--Wyman--Changeaux model discussed below.

To continue, we need to count how many signaling molecules are bound to the receptor, and then changes in this number act as a source or sink for diffusion.  The total number of bound ligands is $n_{\rm total} = \sum_{\rm i} n_{\rm i} p_{\rm i}$, and hence the diffusion equation for the signaling molecule becomes
\begin{equation}
{{\partial c({\bf x}, t)}\over{\partial t}}
= D \nabla^2 c({\bf x} , t) - \delta ({\bf x} - {\bf x}_0)\sum_{\rm i} n_{\rm i} {{dp_{\rm i}}\over {dt}}.
\label{diffusion1}
\end{equation}
We can solve explicitly for the response to small changes in the populations, again working in the   frequency domain:
\begin{equation}
\delta{\tilde c}({\bf x}; \omega ) = -\int {{d^3k}\over {(2\pi )^3}}
{{e^{i{\bf k \cdot}({\bf x} - {\bf x}_0)}}
\over
{-i\omega + Dk^2}}
(-i\omega ) \sum_{\rm i} n_{\rm i} \delta{\tilde p}_{\rm i}(\omega)
\end{equation}
In particular, to substitute into Eq (\ref{chi2}) we need to evaluate the concentration at the location of the receptor itself
\begin{equation}
\delta{\tilde c}({\bf x}_0; \omega )
= 
i\omega \left[\int {{d^3k}\over {(2\pi )^3}}
{1
\over
{-i\omega + Dk^2}}\right]
 \sum_{\rm i} n_{\rm i} \delta{\tilde p}_{\rm i}(\omega)
\end{equation}
The term in brackets is ultraviolet divergent  because we treat the receptor as a point object.  As in Ref \cite{pnas} we remove the divergence by cutting off the $k$ integrals at a scale $k_{\rm max} \sim \pi/\ell$, where $\ell$ is the size of the receptor site.  Then  we find, in  the low frequency limit $\omega \ll D/\ell^2$,
\begin{equation}
\delta{\tilde c}({\bf x}_0; \omega \rightarrow 0)
=
{{i\omega }\over{2\pi D\ell}}
 \sum_{\rm i} n_{\rm i} \delta{\tilde p}_{\rm i}(\omega) .
 \label{c_from_p}
\end{equation}

Equation (\ref{c_from_p}) tells us how the concentration responds to changes in the population of the different states of the receptor complex, while Eq (\ref{chi2}) tells us how the populations respond to changes in concentration.  In effect the diffusing concentration field keeps a history of the state dynamics, and therefore  renormalizes the kinetics, as shown explicitly for a simpler case in Ref \cite{pnas}.  To finish the calculation we substitute from Eq (\ref{c_from_p}) into (\ref{chi2}) and invert the matrix relations to obtain
\begin{equation}
\delta {\tilde E}_{\rm i}(\omega ) = \sum_{\rm j} 
\left[
\chi^{-1}_{\rm ij} (\omega ) 
+ n_{\rm i} n_{\rm j} {{i\omega k_B T}\over{2\pi D \ell {\bar c}}}
\right]
\delta {\tilde p}_{\rm j} (\omega )   ,
\end{equation}
where  $\chi^{-1}_{\rm ij}$ denotes the $\rm ij$ element of the matrix inverse to $\chi_{\rm ij}$.

We see that the effect of coupling to diffusion is to add  a self--energy term to the inverse susceptibility, so that
\begin{equation}
\chi_{\rm ij}^{-1} (\omega ) \rightarrow \chi_{\rm ij}^{-1} (\omega ) + n_{\rm i} n_{\rm j}
{{i\omega k_B T}\over{2\pi D \ell {\bar c}}} .
\end{equation}
The fluctuation--dissipation theorem tells us that the fluctuations in occupancy of the states can be thought of as the response to fluctuations in the conjugate energies $E_{\rm i}$, and the spectrum of these fluctuations is determined by the inverse susceptibility \cite{signs}:
\begin{equation}
\langle \delta {\tilde E}_{\rm i} (\omega )  \delta {\tilde E}_{\rm j} (\omega ' ) \rangle
= 2\pi\delta (\omega + \omega ') {{2k_B T}\over{\omega}} {\rm Im} \left[\chi_{\rm ij}^{-1}(\omega ) \right]
\label{fdt}
\end{equation}
If we define the noise power spectrum through
\begin{equation}
\langle \delta {\tilde E}_{\rm i} (\omega )  \delta {\tilde E}_{\rm j} (\omega ' ) \rangle
= 2\pi\delta (\omega + \omega ') {\cal N}_{\rm ij} (\omega ),
\end{equation}
\begin{equation}
 {\cal N}_{\rm ij} (\omega ) =  {\cal N}_{\rm ij}^{(0)} (\omega ) + 
 n_{\rm i} n_{\rm j} {{(k_B T)^2}\over{\pi D\ell {\bar c}}} ,
 \label{calNfinal}
\end{equation}
where ${\cal N}_{\rm ij}^{(0)}$ is the noise we would calculate just from the kinetics described by the bare susceptibility $\chi_{\rm ij}$.

We see from Eq (\ref{chi2}) that a change  in concentration is equivalent to a coordinated change in the energies of all the states,
\begin{equation}
\Delta {\tilde E}_{\rm i} (\omega ) = -n_{\rm i} k_B T {{\Delta {\tilde c({\bf x}_0; \omega )}}\over{\bar c}} .
\end{equation}
Because the change in concentration produces changes in the free energies of many different states, in general there is no single state of the receptor complex that can be monitored to provide the optimal readout of the concentration.  If we imagine, however, that downstream mechanisms have access to all the states, and can form a readout constructed as a weighted average of the populations, then by choosing the weights appropriately one can achieve the maximum possible signal to noise ratio at each frequency \cite{threshold},
\begin{eqnarray}
SNR(\omega ) &=& \sum_{\rm ij} \Delta {\tilde E}_{\rm i} (\omega ) {\cal N}_{\rm ij}^{-1}
(\omega ) \Delta {\tilde E}_{\rm j}^* (\omega )\\
&=& {\bigg |} {\Delta {\tilde c({\bf x}_0; \omega )}\over{\bar c}} k_B T {\bigg |}^2
 \sum_{\rm ij} n_{\rm i}  {\cal N}_{\rm ij}^{-1} (\omega )
n_{\rm j} 
\end{eqnarray}
In the limit that we are trying to detect a slow change in concentration, and we are willing to average for a time $\tau_{\rm avg}$, the total signal--to--noise ratio is given by $\tau_{\rm avg} SNR(\omega \rightarrow 0)$.
Defining the detectability as the point where 
 this total signal--to--noise ratio reaches unity,  the threshold for detection is
\begin{equation}
{{\Delta c}\over {\bar c}} = \left[
\tau_{\rm avg}  (k_B T)^2 \sum_{\rm ij} n_{\rm i}  {\cal N}_{\rm ij}^{-1} (\omega =0 )
n_{\rm j} 
\right]^{-1/2}
\label{dc1}
\end{equation}

The sum in Eq (\ref{dc1}) is a quadratic form, ${\bf n}^{T} {\bf \cdot} {\cal N}^{-1} {\bf \cdot n}$, where the vector $\bf n$ has components $n_{\rm i}$.  
If we define $\bf u$ to be  unit vector parallel to $\bf n$, then we can write
\begin{equation}
{\cal N} = ({\bf u}^T {\bf \cdot} {\cal N}{\bf \cdot u}) {\bf u u}^T + {\cal N}_{\perp},
\end{equation}
where the matrix ${\cal N}_{\perp}$ acts only in the subspace orthogonal to $\bf n$, and then
${\bf n}^{T} {\bf \cdot} {\cal N}^{-1} {\bf \cdot n} = {{{\big |}{\bf n}{\big |}^2}/
({{\bf u}^T {\bf \cdot} {\cal N}{\bf \cdot u}}})$ .
From Eq (\ref{calNfinal}),
\begin{equation}
{\cal N} = {\cal N}^{(0)} + {{(k_B T)^2}\over{\pi D\ell {\bar c}}} {\bf n n}^T ,
\end{equation}
so that 
\begin{equation}
{\bf u}^T {\bf \cdot} {\cal N}{\bf \cdot u} = {\bf u}^T {\bf \cdot} {\cal N}^{(0)}{\bf \cdot u}
+ {{(k_B T)^2}\over{\pi D\ell {\bar c}}}{\big |}{\bf n}{\big |}^2 .
\end{equation}
Putting all of the factors together, we have
\begin{equation}
{{\Delta c}\over {\bar c}} =
\left[
{1\over {\pi D\ell {\bar c}\tau_{\rm avg}}} + {{{\bf u}^T {\bf \cdot} {\cal N}^{(0)} {\bf \cdot u}}
\over {{\big |}{\bf n}{\big |}^2 \tau_{\rm avg} (k_B T)^2}}
\right]^{1/2}
\end{equation}
Crucially, the second term in brackets is positive, because it is a quadratic form constructed from the matrix of noise power spectra.  Thus we have, exactly as in Ref \cite{pnas},
\begin{equation}
\left( {{\Delta c}\over {\bar c}}\right)^2  =
{1\over {\pi D\ell {\bar c}\tau_{\rm avg}}} + {\rm stuff},
\label{limit_with_stuff}
\end{equation}
where $\rm stuff$ is positive and hence the first term provides a lower bound the smallest detectable signal.  This term is, except for the factor of $\pi$, exactly the expression in Eq (\ref{bp}), derived heuristically by Berg and Purcell \cite{BP}.  

\begin{figure}
 \epsfxsize=2.25 in
\epsffile{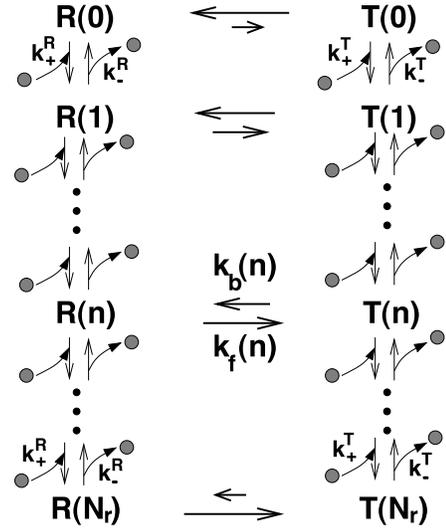}
\caption{The Monod--Wyman--Changeux model for cooperative binding \cite{mwc}, shown here as used in \cite{alon+al_98} to describe the bacterial flagellar motor.
There are $N_r$ sites at which the ligand can bind, and all binding events are independent, but the binding energy depends on whether the whole complex is in the $R$ or $T$ state.  In this case the binding is stronger in the $T$ state, so that by detailed balance the equilibrium shifts toward $T$ (shown schematically by the horizontal arrows) as more ligands bind.}
\label{mwc_fig}
\end{figure}

To illustrate these ideas, it is useful to think about a specific model, the Monod--Wyman--Changeux model of cooperativity \cite{mwc}, schematized in Fig \ref{mwc_fig}.  The MWC model has been widely used, first for allosteric enzymes, then for the paradigmatic example of cooperative oxygen binding to hemoglobin \cite{perutz_90}, and more recently for the analysis of the clockwose/counterclockwise switching of the bacterial flagellar motor \cite{scharf+al_98, turner+al_99,alon+al_98}.
In this model, ligands bind independently to multiple sites, but the binding energy depends on whether the whole complex is in the $R$ or $T$ state \cite{r/t}.  

Quantitatively, the states of the system are defined by the binary variable $R/T$ and the number $n$ of ligands bound to a total of $N_r$ sites.  We can write the free energy of the $R$ state with $n$ ligands bound as
\begin{equation}
F_R(n) = F_R(0) - n k_B T\ln\left({c\over{K_R}}\right) ,
\end{equation}
and similarly for the $T$ states, where $c$ is the ligand concentration as above.  If the rate for ligand binding to the $R$ state [i.e., for the transition $(R,n)\rightarrow (R,n+1)$] is $k_+^R c$ and the rate of unbinding is $k_-^R$, then $k_-^R / k_+^R = K_R$.  If the transition $(R,n) \rightarrow (T,n)$ occurs at rate $k_f(n)$, and the reverse $T\rightarrow R$ transition at rate $k_b(n)$, then by detailed balance around the loops in Fig \ref{mwc_fig} we must have 
\begin{equation}
{{k_f(n)}\over{k_b(n)}} = {{k_f(0)}\over{k_b(0)}} \left( {{K_R}\over{K_T}}\right)^n .
\end{equation}
To complete the kinetic model we follow Ref \cite{turner+al_99} and assume that the activation energies for the transition rates $k_{f,b}(n)$ are themselves linear in $n$, so that
\begin{eqnarray}
k_f(n) &=& k_f(0) \left( {{K_R}\over{K_T}}\right)^{\alpha n}
\label{kfn}\\
k_b(n) &=& k_b (0)\left( {{K_R}\over{K_T}}\right)^{(\alpha-1) n} .
\end{eqnarray}

In this model the equilibrium probability of being in the $T$ state (summing over $n$) is given by
\begin{equation}
\bar p_T = \left[ 1 + {1\over L} \left( {{1 + c/K_R}\over{1+c/K_T}}\right)^{N_r}\right]^{-1} ,
\end{equation}
where $L = \exp\left[ - (F_T(0) - F_R(0))/k_B T\right]$.  In the limit that there are very many binding sites ($N_r\rightarrow \infty$) and binding to the $T$ state is much stronger than to the $R$ state ($K_R \gg K_T$), 
$p_T$ approaches a step function dependence on concentration.
In this sense, the MWC model  allows for arbitrarily high sensitivity if there are enough binding sites that can cooperate.  

In the (plausible) limit where the conformational transition is slow but the binding and unbinding of the ligands is fast,  the dynamics reduces to a two--state system ($R$ and $T$) and we can compute the transition rates as averages of the $k_{f,b}$ over the equilibrium distribution of $n$ given that the system is in either the $R$ or $T$ state.  Once we have this effective two state system, we have to couple the transitions to diffusion of the ligand as before, and then follow the same steps through the fluctuation--dissipation theorem to the effective noise level for concentration  measurements, as in Ref \cite{pnas}.  We outline the argument here and give details elsewhere.

Given that the system is in the $R$ state, the equilibrium distribution of the number of ligand molecules bound is  binomial  because the binding events are independent:
\begin{equation}
p(n|R) = \left(\begin{array}{c}N_r\\n \end{array} \right)  f_R^n (1-f_R )^{N_r - n} ,
\end{equation}
where the fractional occupancy is $f_R = c/(c+K_R)$, and there is an identical equation for the $T$ state.  Then the average rate $\bar k_f$ for transitions from $R$ to $T$ is given by
\begin{eqnarray}
{\bar k}_f &=& \sum_{n=0}^{N_r} p(n|R) k_f(n)\\
&=& k_f (0) \left[ {{K_R + c(K_R /K_T )^\alpha }\over{K_R + c}}\right]^{N_r} ,
\label{kfbar}
\end{eqnarray}
where we use $k_f(n)$ from Eq (\ref{kfn}).
The same argument for the $T$ state yields
\begin{equation}
{\bar k}_b = k_b (0) \left[ {{K_T + c(K_R /K_T )^{1-\alpha} }\over{K_T + c}}\right]^{N_r} ,
\label{kbbar}
\end{equation}
and the dynamics of the $T$ state population is given by
\begin{equation}
{{dp_T}\over{dt}} = {\bar k}_f (1-p_T) - {\bar k}_b p_T .
\label{pTdynamics}
\end{equation}
In the $R$ state the mean number of ligands bound is just
$\bar n_R = N_r f_R$ and similarly in the $T$ state, so that transitions from the $T$ to $R$ state are associated with the release of $N_r (f_T - f_R)$ molecules into the surrounding solution.  Hence the diffusion equation analogous to Eq (\ref{diffusion1}) becomes
\begin{equation}
{{\partial c({\bf x}, t)}\over{\partial t}}
= D \nabla^2 c({\bf x} , t) - \delta ({\bf x} - {\bf x}_0)N_r (f_T - f_R) {{dp_T}\over{dt}}.
\label{diffusion2}
\end{equation}

Solving Eq (\ref{diffusion2}) to compute the response of $c$ to small changes in $p_T$ follows the same steps as for Eq (\ref{diffusion1}), with the result that
\begin{equation}
\delta \tilde c({\bf x}_0; \omega) = {{i\omega}\over{2\pi D\ell}} N_r(f_R - f_T) \delta\tilde p_T(\omega ) .
\label{dc_mwc}
\end{equation}
  The analysis of Eq (\ref{pTdynamics}) is more subtle because the concentration $c$ and the thermodynamic force conjugate to $p_T$ are hidden in the rate constants $\bar k_{f,b}$.  The conjugate force is the free energy difference  between the $T$ and $R$ states, and is linked to the rate constants through detailed balance.  Thus if we imagine changing the rate constants $\bar k_{f,b}$ by small amounts $\delta \bar k_{f,b}$, we must have (see also Ref \cite{pnas})
\begin{equation}
{{\delta \bar k_f}\over{\bar k_f}} - {{\delta\bar k_b}\over{\bar k_b}} = { {\delta F}\over{k_B T}} + N_r (f_T - f_R) {{\delta c}\over {\bar c}} ,
\end{equation}
as can be verified by detailed computations from Eq's (\ref{kfbar}) and (\ref{kbbar}), identifying $F = F_R(0) - F_T(0)$.
Linearizing Eq (\ref{pTdynamics}) thus yields
\begin{widetext}
\begin{eqnarray}
{{d\delta p_T}\over{dt}}
&=& - (\bar k_f + \bar k_b) \delta p_T + \delta\bar k_f (1-\bar p_T) - \delta \bar k_b \bar p_T
\\
&=&
- (\bar k_f + \bar k_b)\delta p_T 
+ {{\delta \bar k_f}\over{\bar k_f}} [\bar k_f (1-\bar p_T)]
- {{\delta\bar k_b}\over{\bar k_b}} [\bar k_b \bar p_T]\\
&=&
- (\bar k_f + \bar k_b)\delta p_T 
+ [\bar k_f (1-\bar p_T)]
\left[{{\delta \bar k_f}\over{\bar k_f}} - {{\delta\bar k_b}\over{\bar k_b}}\right]\\
&=& - (\bar k_f + \bar k_b)\delta p_T 
+ [\bar k_f (1-\bar p_T)]\left[ { {\delta F}\over{k_B T}} + N_r (f_T - f_R) {{\delta c}\over {\bar c}} \right] ,
\label{linearized_pT}
\end{eqnarray}
where we have used the fact that the equilibrium occupancy of the $T$ state satisfies 
$\bar k_f (1-\bar p_T) = \bar k_b p_T$, corresponding to $dp_T/dt = 0$ in Eq (\ref{pTdynamics}). Fractional changes in concentration are equivalent to changes in free energy through
\begin{equation}
{{\delta c}\over c} = {1\over{N_r (f_T - f_R)}} {{\delta F}\over{k_B T}} ,
\label{cequiv}
\end{equation}
so if we can compute the noise in $\delta F$ we can compute the equivalent noise in the concentration, as before.  

To find the susceptibility of $p_T$ to its conjugate force $\delta F$ we transform Eq (\ref{linearized_pT}) and substitute from Eq (\ref{dc_mwc}) to give
\begin{equation}
{{\delta \tilde F(\omega )}\over{\delta \tilde p_T(\omega )}}
= {{k_B T}\over{(1-\bar p_T)}} (\bar k_f + \bar k_b ) 
-i\omega\left[
{{k_B T}\over{\bar k_f (1-\bar p_T)}}  + {k_B T{N_r^2 (f_T - f_R)^2}\over{2\pi D\ell \bar c}}\right] .
\end{equation}
The fluctuation--dissipation theorem, as in Eq (\ref{fdt}) but now with the more standard sign convention \cite{signs}, tells us that
\begin{equation}
\langle \delta \tilde F(\omega ) \delta \tilde F(\omega ')\rangle = -{{2k_B T}\over\omega}
{\rm Im}\left[ {{\delta \tilde F(\omega )}\over{\delta \tilde p_T(\omega )}} \right] .
\end{equation}
By analogy with the arguments above this result, combined with Eq (\ref{cequiv}) means that the equivalent concentration noise level is
\begin{eqnarray}
\left( {{\Delta c}\over {\bar c}}\right)^2  &=& {1\over {\tau_{\rm avg}}}\left[ {1\over{N_r (f_T - f_R)(k_B T)}} \right]^2
\left[-{{2k_B T}\over\omega}\right]{\rm Im}\left[ {{\delta \tilde F(\omega )}\over{\delta \tilde p_T(\omega )}} \right]\\
&=& {1\over{\pi D \ell \bar c \tau_{\rm avg}}} + {2\over{N_r^2 (f_T - f_R)^2}} \cdot {1\over {\bar k_f \tau_{\rm avg} (1-\bar p_T)}} .
\label{mwc_result}
\end{eqnarray}
\end{widetext}
We see that this is exactly of the form found in our more general analysis, except that now we can give the $\rm stuff$ of Eq (\ref{limit_with_stuff}) an explicit form.  In particular, as the number of cooperative sites $N_r$ becomes large, this extra term becomes small, and the physical limit set by diffusion alone becomes more dominant.


To summarize, we have found that the physical limit to biochemical signaling first suggested by Berg and Purcell is surprisingly general.  Even allowing for arbitrarily complex internal states and multiple ligand binding sites, the equivalent noise level against which concentration changes must be detected has two terms:  the Berg--Purcell limit plus a positive contribution from the details of the chemical kinetics.  Cooperative interactions serve to suppress the second term, and perhaps this is crucial in allowing any real biological system to approach the physical limit.
We conclude by recalling another of the heuristic arguments offered by Berg and Purcell \cite{BP}.  They suggested that one should view chemical sensing as molecule counting, and then by analogy with photon counting the physical limit is set by shot noise in the flux; the difference is that for molecules the flux is set by diffusion.  Our results are completely consistent with this point of view: we can think of the detailed kinetics of the system as ``processing'' the input flux, and then noise can be added in this processing but of course the shot noise cannot be reduced.  Cooperativity provides for a high gain in response to the input and hence reduces the impact of extraneous noise sources.

\acknowledgments{Work at Princeton was supported in part by the Materials Research Science and Engineering Centers Program of the National Science Foundation under Award DMR--0213706, by National Institutes of Health Grant P50 GM071508, and by the Princeton University Science and Technology Council.}


\begin{thebibliography}{99}
\bibitem{signs}
The sign of the noise spectrum here is opposite to what one finds in textbooks.  The reason is that the energy of the system is $+\sum_{\rm i} p_{\rm i} E_{\rm i}$, whereas when we think about forces and displacements we have the energy $-\sum_{\rm i} x_{\rm i} f_{\rm i}$.
\bibitem{r/t}
In the original discussion \cite{mwc}, $R$ referred to a ``relaxed'' state of the molecule and $T$ to a ``tense'' state.  In Ref \cite{alon+al_98} and Fig \ref{mwc_fig}, $R$ and $T$ refer to running and tumbling motions that result from the different directions of rotation of the flagellar motor.  The notational shift is important, because in the original work the ligand bound more strongly to the $R$ state, but in the later discussion it is the $T$ state that has higher affinity.
\end{thebibliography}
 \end{document}